# The Properties of an Hɪ-Selected Galaxy Sample

Arpad Szomoru[1], Puragra Guhathakurta[2,3], Jacqueline H. van Gorkom[4],
Johan H. Knapen[5], David H. Weinberg[6], and Andrew S. Fruchter[7,3]

[1] Kapteyn Astronomical Institute, P.O. Box 800, NL 9700 AV Groningen, The Netherlands
[2] Hubble fellow, Princeton University Obs., Peyton Hall, Princeton, NJ 08544, USA
[3] Current address: STScI, 3700 San Martin Drive, Baltimore, MD 21218, USA
[4] Department of Astronomy, Columbia University, New York, NY 10027, USA
[5] Instituto de Astrofísica de Canarias, E-38200 La Laguna, Tenerife, Spain
[6] Institute for Advanced Study, Princeton, NJ 08540, USA
[7] Hubble fellow, Astronomy Dept., University of California, Berkeley, CA 94720, USA

**Abstract**

We analyze the properties of a sample of galaxies identified in a 21-cm, Hɪ-line survey of selected areas in the Perseus-Pisces supercluster and its foreground void. Twelve fields were observed in the supercluster, five of them (target fields) centered on optically bright galaxies, and the other seven (blank fields) selected to contain no bright galaxies within $45'$ of their centers. We detected nine previously uncataloged, gas-rich galaxies, six of them in the target fields. We also detected Hɪ from seven previously cataloged galaxies in these fields. Observations in the void covered the same volume as the twelve supercluster fields at the same Hɪ-mass sensitivity, but no objects were detected.

Combining our Hɪ data with optical broad-band and H$\alpha$ imaging, we conclude that the properties of Hɪ-selected galaxies do not differ substantially from those of late-type galaxies found in optical surveys. In particular, the galaxies in our sample do not appear to be unusually faint for their Hɪ mass, or for their circular velocity. We find tentative evidence for a connection between optical surface brightness and degree of isolation, in the sense that low surface brightness galaxies tend to be more isolated. The previously cataloged, optically bright galaxies in our survey volume dominate the total Hɪ mass density and cross-section; the uncataloged galaxies contribute only $\sim 19\%$ of the mass and $\sim 12\%$ of the cross-section. Thus, existing estimates of the density and cross-section of neutral hydrogen, most of which are based on optically-selected galaxy samples, are probably accurate. Such estimates can be used to compare the nearby universe to the high-redshift universe probed by quasar absorption lines.

## 1 Introduction

Most investigations of the properties or the spatial distribution of galaxies begin with a sample selected in the optical, or sometimes in the infrared (based on *IRAS* [Neugebauer et al. 1984] 100 $\mu$m detections). Even 21-cm or radio continuum studies usually target a sample of galaxies taken from optical catalogs, like the Uppsala Galaxy Catalog (UGC; Nilson 1973) or the Catalog of Galaxies and Clusters of Galaxies (Zwicky et al. 1961–1968). Thus, any population of galaxies or gas clouds with very low optical luminosity or surface brightness could have largely escaped detection and study.

To circumvent this problem, we have conducted a direct search in the 21-cm line of neutral hydrogen, using the D-configuration of the Very Large Array (VLA[1]; Napier et al. 1983), covering fields in the Perseus-Pisces supercluster and in its foreground void (see Haynes & Giovanelli 1989, hereafter HG). In an earlier paper (Weinberg et al. 1991, hereafter Paper I), we summarized the HI properties of our detected objects, and we discussed the implications of our results for the shape of the HI mass function and for the spatial distribution of dwarf galaxies. In this paper, we describe the 21-cm observations and results in greater detail, and we also present the results of follow-up optical observations, both broad-band and H$\alpha$ imaging. These data allow us to compare directly the properties of our HI-selected objects to those of optically selected galaxies.

For the VLA survey, we observed twelve fields in the Perseus-Pisces supercluster and thirty fields in the foreground void. The effective survey volume varies with HI mass, since more massive objects can be detected farther from the center of the primary beam and at greater distances from the earth. Paper I discussed the resulting "selection function" of the survey and compared it to that of previous 21-cm searches, particularly that of Fisher & Tully (1981), which at the time was the largest directly comparable survey. We will not repeat the details here—the salient points are that our survey covered nearly equal volumes in the void and the supercluster, and that its total effective volume, roughly 100 Mpc$^3$ for $M_{\rm HI} > 3 \times 10^8\, M_\odot$ and 200 Mpc$^3$ for $M_{\rm HI} > 10^9\, M_\odot$, was much larger than that of previous surveys for HI masses in the $10^8$–$10^9\, M_\odot$ range typical of gas-rich dwarf galaxies.[2] Henning (1992) has recently published results of a much larger 21-cm survey, which covers regions that are obscured by the Galactic plane at optical wavelengths. The sensitivity of these observations varied from one line-of-sight to another because of solar and man-made interference, dependence on declination and frequency resolution (Henning 1992); we are therefore unable to make a quantitative comparison of survey volumes. The survey detected 37 extragalactic objects, 19 of them previously cataloged. Follow-up optical analysis of the sample is still in progress. For a review of other 21-cm surveys, see Briggs (1990).

Our survey detected nine previously uncataloged galaxies, along with seven brighter galaxies that are listed in the UGC or the Zwicky catalog. Our definition of supercluster and void fields is based on HG's redshift surveys of optically selected galaxies. This strategy allows us to address some questions of large-scale structure even with a small survey volume, but it complicates the interpretation of our results to some extent. In particular, it is unclear whether the optically cataloged galaxies should be considered part of our HI-selected "sample", since they played a role in the choice of survey fields. Our approach in this paper is to present results for our full sample but to focus on the previously uncataloged objects as a distinct subset. All of the uncataloged objects are fainter (in $B$, $R$, and $I$) than all of the previously cataloged galaxies, and for convenience we will hereafter refer to them as "H I-dwarfs". However, only three of the nine have absolute $B$-magnitudes fainter than $-16$, which is the boundary often used in conventional, optical definitions of dwarf galaxies.

---

[1] The VLA of the National Radio Astronomy Observatory (NRAO) is operated by Associated Universities, Inc., under a cooperative agreement with the National Science Foundation.

[2] Unless stated otherwise, we have adopted a Hubble constant $H_0 = 100\,{\rm km\,s^{-1}\,Mpc^{-1}}$ in this paper.

The results of our survey, combined with the follow-up optical data, allow us to examine both the properties and the spatial distribution of H I-selected galaxies. We are limited by small numbers in both cases, but because we are sensitive to objects that could have been entirely missed by earlier studies, even results with large statistical uncertainties are important; they provide a rough guide to previously unexplored territory. Although we covered equal volumes in the void and supercluster, all our detections were in supercluster fields. As discussed in Paper I, this result implies that the H I-dwarfs trace the structure outlined by optically bright galaxies, at least qualitatively, though we cannot rule out a modest "bias" between the H I-dwarfs and the giants. Our other significant finding in Paper I was that the number of objects detected was no more than one would predict by combining a standard estimate of the optical luminosity function with the H I-mass-to-optical-luminosity relation that describes optically selected dwarf galaxies. From this statistical evidence, we inferred that our 21-cm search had not turned up a new population of gas-rich, optically faint objects, but that it had rather detected the "normal" population of irregular galaxies. The optical data in this paper allow us to address this point more directly.

We describe our observing procedures in the next section, present our results in §3, and discuss their implications in §4.

## 2  Observations and Data Analysis

### 2.1  21 cm Spectral Line

The VLA spectral line observations took place during November 1989. The parameters of the observations are listed in Table 1. All 27 antennae were employed in D-configuration resulting in an angular resolution of about $1'$ (full width at half maximum [FWHM] of the synthesized beam). The bandwidth of the 21-cm observations was 6.25 MHz, corresponding to a usable velocity range of about 1200 km s$^{-1}$. On-line Hannning smoothing was used after which every other channel was discarded, resulting in sets of 31 independent channels and a velocity resolution of $\sim$42 km s$^{-1}$. The total observing time, including time spent on phase and flux calibration, was 70 hr.

All the VLA pointings were located within the strip $0^h0^m0^s < \alpha_{1950} < 1^h57^m36^s$ and $30° < \delta_{1950} < 34°$. The pointings are listed in Table 2. For each field we made $1° \times 1°$ images, mapping out the full extent of the primary beam. Twelve fields were observed in the Perseus-Pisces supercluster at a redshifted velocity of 5100 km s$^{-1}$, and 30 fields in the foreground void at 3250 km s$^{-1}$. The volume surveyed in a single pointing scales as $D^2$, where $D$ is the distance corresponding to the mean velocity to which the VLA bandpass is tuned; the number of supercluster and void fields were chosen so that the total volume surveyed in each region was roughly the same. The integration times $t_{\rm int}$ were 40 min for each of the void fields and 210 min for each of the supercluster fields, roughly in the ratio of $D^4$ to ensure equal H I-mass sensitivity in the void and supercluster ($M_{\rm HI} \propto \sigma D^2$ and $\sigma \propto t_{\rm int}^{-1/2}$, where $\sigma$ is the r.m.s. noise in the 21-cm data).

Five of the 12 supercluster fields were centered on Perseus-Pisces galaxies brighter than $M_B = -18.4$ (target fields). These optically bright galaxies were chosen from the redshift catalog of HG. The remaining seven supercluster fields were selected to have no bright galaxies within $45'$ of their pointing centers (blank fields).

The UV data were calibrated and processed into ($\alpha$, $\delta$, $V$) datacubes using standard techniques in NRAO's Astronomical Image Processing System. To achieve the best possible signal-to-noise ratio, natural weighting was applied in making the cubes. Subsequent analysis of the VLA datacubes was carried out using the Groningen Image Processing System (GIPSY). We searched for H I line emission by subtracting the 21-cm continuum emission from each cube; preliminary estimates of the continuum map were obtained by (1) averaging channels 4–28 of the datacube, and (2) averaging channels 4–8 and 24–28. If line emission was detected, the continuum subtraction was redone using the average of the line-free channels on the low and high frequency side of the line emission. In a few cases, the continuum was subtracted by fitting a linear baseline to the line-free channels. Images with H I emission were CLEANed and restored with a gaussian beam. The r.m.s. noise $\sigma = 1$ mJy/beam in the final images of the void fields and 0.4 mJy/beam in the supercluster fields.

Total H I images were constructed by blanking areas without line emission and summing the channel maps. Neutral hydrogen masses were calculated using the formula:

$$M_{\rm HI}/M_\odot = 2.36 \times 10^5 \, D^2 \int S \, dV$$

where $D$ is the distance in Mpc, and the flux density $S$ (in Jy) is integrated over the 21-cm line, with the velocity $V$ expressed in km s$^{-1}$. The mean heliocentric velocity $V_{\rm hel}$ of each galaxy was determined from its global H I line profile by averaging the velocities at which the flux density falls to 20% and 50% of the peak value on each side of the 21-cm line. We compute the distance to each galaxy by correcting its heliocentric velocity for the motion of the sun with respect to the velocity centroid of the Local Group: $D = [V_{\rm hel} + 300\sin(l)\cos(b)]/H_0$, where $l$ and $b$ are the Galactic longitude and latitude of the galaxy, and $H_0$ is the Hubble constant.

The observed line widths at 20% and 50% of the peak value, $W_{20}$ and $W_{50}$, are listed in Table 3. Also listed is $\log(W_{\rm R}^i)$, which is the logarithm of $W_{20}$, corrected for instrumental broadening following Bottinelli et al. (1990), for turbulent broadening following Tully & Fouqué (1985) and corrected to edge-on. The magnitude of the instrumental broadening correction, which is based on a statistical intercomparison of line profiles measured with various velocity resolutions, is $-23.4$ km s$^{-1}$.

The majority of the uncataloged galaxies is barely resolved in H I. An estimate of their angular size was obtained by fitting a two-dimensional gaussian to the total H I distribution; the sizes quoted for the H I-dwarfs are the FWHM of these gaussians with the FWHM of the beam subtracted in quadrature. For the bright galaxies, the H I column density contour at $5 \times 10^{19}$ cm$^{-2}$ was used to define the angular size. It should be kept in mind, however, that the size determinations for the smaller galaxies are only rough indications of the true sizes because of the marginal angular resolution. Intensity-weighted first moment maps were used to construct velocity fields for the large galaxies in which the H I distribution was well resolved both spatially and in velocity.

## 2.2 Optical Broad-band

Optical observations of the galaxies detected in H I were carried out during the second week of October 1990 using the 1-m Jacobus Kapteyn Telescope (JKT) on the island of La Palma. The faintest of the galaxies are uncataloged and are barely visible on the Palomar Optical Sky Survey plates. The camera on the JKT consisted of a GEC charge-coupled device (CCD) with $385 \times 578$ pixels and a pixel scale of $0''.3$. Broad-band optical images were obtained through $B$ (4500 Å), $R$ (6500 Å), and $I$ (9000 Å) filters. Integration times were $2 \times 900$ s in $B$ ($3 \times 900$ s for the faintest objects), 900 s in $R$, and $2 \times 900$ s in $I$. The seeing ranged from $0''.9$ to $2''$, and the sky conditions were photometric during most of the observations. We obtained short exposures of bright standard stars (Landolt 1983) at regular intervals throughout the photometric nights, and we used these to photometrically calibrate the JKT images. All the galaxies were observed close to transit so that airmass corrections were minimal. The estimated errors in the magnitude zero points are $B = 0.05$ mag, $R = 0.04$ mag and $I = 0.04$ mag. Twilight sky flats were used for flat fielding the CCD images. Sky subtraction was done by fitting first- and second-order polynomial surfaces to blank regions of sky around the image of each galaxy.

Three of the galaxies, UGC 598b, PPHI 0023+34, and PPHI 0147+31, are too faint to be detected in the JKT $B$ images. These objects were reobserved in service time with the 4.2-m William Herschel Telescope (WHT) in November 1990 using the Taurus instrument in imaging mode. A $1180 \times 1280$ EEV CCD was employed, with a pixel size of $0''.27$. One 600 s exposure was obtained for each of the three galaxies through a $B$ filter. Twilight sky flats were used for flat fielding the CCD frames. An artifact in the sky flat image affected the flat fielding of the UGC 598b image; the flux determination of this galaxy is therefore uncertain. After sky subtraction, the WHT frames were calibrated by a bootstrap procedure: the fluxes of some bright stars in the WHT images were compared with the photometrically calibrated fluxes of the same stars in the JKT $B$ images.

The total magnitude of each galaxy was calculated by integrating over a circular aperture of radius $30''$, except for UGC 26 for which an aperture of radius $50''$ was used. In a few cases, images of foreground stars superposed on the galaxy images were removed; imperfect star subtraction is an additional source of uncertainty in the flux measurement of these galaxies. Magnitudes were corrected for absorption by dust along the line-of-sight in the Galaxy following the precepts of Burstein & Heiles (1984). For comparison with existing samples, we have also applied standard corrections for internal absorption as a function of wavelength and disk inclination (Tully & Fouqué 1985).

In order to derive surface photometry of the galaxies, we fitted ellipses to a succession of isophotes of the $I$-band light distribution (of decreasing brightness and increasing radius). Radial brightness profiles in the $B$, $R$, and $I$ bands were constructed by integrating the light in these images within the best-fit ellipses. The brightness profiles of the galaxies are well approximated by exponential laws and these were used to obtain estimates of the central surface brightness $\mu_0$, the disk scale length $h$, the diameter $D_{25}$ of the $B = 25$ mag arcsec$^{-2}$ isophote, and the $B - R$ and $R - I$ color as a function of radius. The best-fit ellipses also yielded an axis ratio and a position angle for each galaxy. Disk inclinations were derived from the isophotal axis ratios by accounting for the finite thickness of stellar disks, assuming an intrinsic axis ratio of 0.2 and correcting for the 3° measurement bias (Tully 1988).

## 2.3 Narrow-band H$\alpha$

The narrow-band H$\alpha$ imaging observations were done on the Shane 3-m telescope at Lick Observatory using the UV Schmidt Spectrograph. Due to limited observing time, only four objects were observed in H$\alpha$: UGC 26a, UGC 26b, UGC 60b, and UGC 598a. Line flux was detected in all four galaxies.

The Cassegrain spectrograph imaging system allowed one to switch between CCD imaging and spectroscopy through the interposition of a flat mirror in the spectroscopic light path. One could therefore use the spectrograph to directly measure the transmission function of the imaging filter, although a small uncertainty remains due to possible spectral dependence of the reflectance of the flat mirror. The H$\alpha$+N II emission of all four galaxies observed fell within the passbands of two Lick narrow-band filters, which have effective wavelengths of 6655 Å and 6695 Å and FWHM of 69 Å and 96 Å, respectively. Continuum (off-band) images of the galaxies were obtained with a 360 Å-wide filter centered at about 7040 Å. Integration times ranged from 2100 s to 3600 s for the H$\alpha$ exposures, and from 1200 s to 2500 s for the continuum exposures. Very light cirrus was present throughout the H$\alpha$ observations, so that our photometry is no better than 5–10%. As a consequence of these non-photometric conditions, we had to determine the appropriate scaling for continuum subtraction on an image-by-image basis by minimizing the residuals of bright stars. We estimate that this empirical method of continuum subtraction provides H$\alpha$+N II fluxes to an accuracy of about 10%.

The measured H$\alpha$+N II fluxes were corrected for Galactic absorption (Burstein & Heiles 1984; Rieke & Lebofsky 1985) and used to estimate star formation rates (SFR) following Hunter & Gallagher (1986). The H$\alpha$ luminosity of a galaxy $L_{\text{H}\alpha}$ was converted to the total number of ionizing photons, which was then compared to the integrated number of photons expected from massive stars for a Salpeter (1955) initial mass function. Extrapolating the Salpeter stellar mass function down to a lower mass limit of 0.1 $M_\odot$, the SFR of all stars (with masses in the range 0.1–100 $M_\odot$) is given by:

$$\dot{M} = 7.07 \times 10^{-42} L_{\text{H}\alpha} \ M_\odot \, \text{yr}^{-1} \quad ,$$

where $L_{\text{H}\alpha}$ is in units of erg s$^{-1}$. Based on this SFR, we calculated the distance-independent quantities $\dot{M}/L_B$ and the timescale $\tau$ over which the current gas supply in a galaxy would last at the current rate of star formation. The gas depletion time $\tau$ is computed by multiplying the H I mass by 1.34, to account for the presence of He, and dividing it by the SFR.

## 3  Results

Sixteen galaxies were detected in H I in our VLA observations of 12 fields in the ridge of the Perseus-Pisces supercluster region. Of these, 13 H I detections are located in 'target' fields centered on optically bright galaxies, while the remaining three are in 'blank' fields away from bright galaxies (see §2.1). Nine of the 16 galaxies detected at 21 cm do not have counterparts in optical catalogs; we refer to these galaxies as "H I-dwarfs" regardless of their optical luminosity. Six of these H I-dwarfs were found in target fields 1–3, and the remaining three in blank fields. Two optically faint, but previously cataloged galaxies, Zw 499.039 and Zw 502.004, were detected in H I in target fields 1 and 4, respectively. One of the targets, the S0a galaxy UGC 598, was not detected in H I, while the fifth target field contains two bright galaxies (UGC 810 and UGC 820). We adopt the following convention: (1) uncataloged galaxies near bright galaxies (*i.e.* in target fields) are denoted by the UGC number of the bright galaxy with "a" or "b" appended, in order of increasing right ascension; and (2) uncataloged galaxies in blank fields are named "PPHI" followed by their 1950 right ascension and declination.

No H I detections were made in the 30 void field observations to a $5\sigma$ flux density limit of 5 mJy/beam, corresponding to an H I mass per channel of $\sim 5 \times 10^7 M_\odot$ at the center of the primary beam. The noise in our datacubes is somewhat non-gaussian because of imperfect subtraction of remote continuum sources; the short duration of the VLA void observations (and the resulting sparse UV coverage) makes it practically impossible to remove these sources exactly.

The H I and radio continuum properties of the galaxies detected in the supercluster are summarized in Table 3. Their optical magnitudes, colors, and morphological properties, based on *BRI* data, are listed in Table 4. Table 5 contains the results of the H$\alpha$ observations, along with other parameters derived from the various datasets, including surface brightness, size, H I-mass-to-light ratio, and gas depletion timescale. Some of the H I parameters in Table 3 (*e.g.* velocity width and total H I mass) differ slightly from those given in Paper I. These differences are a result of redoing a large part of the analysis using the completely rewritten, UNIX-based version of GIPSY; this allowed us to take advantage of several convenient features in the new version and corrected a minor software error in the old package. In nearly every case, the values of the H I parameters have remained equal to the previously published value to within the $1\sigma$ uncertainty. The velocity width of UGC 810 at 20% ($W_{20}$) was listed incorrectly in Table 1 of Paper I, and this has been corrected in Table 3 of this paper.

We present global H I profiles in Figure 1 of all 16 galaxies detected at 21 cm: four galaxies in field 1 (top left panel), three in field 2 (top right panel), two in field 3 (center left), two in field 4 (center right), and two in field 5 (bottom left). The line profiles of the three H I-dwarfs detected in blank fields are shown together in the bottom right panel of Figure 1. Figure 2 shows the velocity fields of six large galaxies, those in which the H I is well resolved in position and velocity. These include the target galaxies of fields 1, 2, 4, and 5 (the target galaxy of field 3, UGC 598, was not detected in H I), along with UGC 810 (the second bright galaxy in field 5) and Zw 502.004 (field 4). Contour plots of the H I column density distributions in the supercluster target fields 1–5 are shown in

Figures 3–7. Figure 8 (four panels) and Figures 9–15 show contour plots of the total H I distribution in some of the individual galaxies superposed on gray-scale representations of their $B$-band images. The shaded ellipses in Figures 2–15 indicate the FWHM of the VLA synthesized beam.

In the rest of this section, the galaxies detected in each of the supercluster fields are discussed in turn.

### 3.1 Field 1

The central galaxy of this field is UGC 26, an SBb galaxy. In addition to UGC 26, three companions were detected around this galaxy in our 21-cm observations, one of which was found to be a cataloged galaxy, Zw 499.039 (Figs. 3 and 8). The four galaxies are at nearly identical redshifts, the largest velocity difference being about $150\,\mathrm{km\,s^{-1}}$ (top left panel of Fig. 1). The distribution of neutral hydrogen in UGC 26 is symmetric (lower left panel of Fig. 8) and its velocity field regular (top left panel of Fig. 2); its double-peaked global profile indicates a disk in differential rotation. The other three galaxies are barely resolved in H I.

Figure 8 shows the H I and blue light distributions of all four galaxies in field 1. The optical surface brightness profiles of UGC 26 deviate significantly from an exponential law due to the effects of a central bar and prominent spiral arms (lower left panel of Fig. 8). As a consequence, the extrapolated central surface brightness and the exponential scale length of the stellar disk are subject to systematic error. In the case of Zw 499.039, a bright foreground star is seen superposed on its disk, and this makes the measurement of its isophotal axis ratio and position angle uncertain. Spiral arms are discernible in the optical images of Zw 499.039 and, as in the case of UGC 26, its disk parameters cannot be determined accurately. The distribution of starlight in the uncataloged galaxy UGC 26a has a very small scale length, and the galaxy appears to be featureless in both optical and H I. Given its very blue $B - R$ color, it could be classified as a compact blue dwarf. The other H I-dwarf in this field, UGC 26b, is slightly asymmetric and has an exponential stellar disk without a clear bulge component.

The H I-dwarfs UGC 26a and UGC 26b were observed in H$\alpha$. The morphology of the line emission (and hence of the ionized gas) in UGC 26a is compact and smooth. By contrast, the H$\alpha$ distribution in UGC 26b is clumpy and extended, indicating that massive star formation is taking place throughout the disk of this galaxy.

### 3.2 Field 2

The VLA observations of this field were centered on the Sb galaxy UGC 60. The target galaxy was detected along with two gas-rich galaxies around it (Fig. 4). One of these, UGC 60a, is close to UGC 60 and may be interacting with it, while the other, UGC 60b, is about $10'$ from the target galaxy. The neutral hydrogen distributions of UGC 60 and UGC 60a overlap partially, with a hint of an H I bridge between them and a faint extension on the opposite side of the larger galaxy (possibly a tidal feature). The velocity fields of both galaxies are shown in the top right panel of Figure 2. The iso-velocity contours of the main galaxy UGC 60 appear to be twisted, suggestive of an interaction. Kinematically, the companion galaxy UGC 60a is distinct from UGC 60, with its major axis at right angles to the major axis of the latter.

We were unable to derive optical surface photometry of the close companion UGC 60a because a bright foreground star happens to lie at the position of the peak of the H I emission. The other H I-dwarf UGC 60b was studied in $BRI$ and in H$\alpha$ (see Fig. 9). Its optical surface brightness profile is exponential with no sign of a bulge component. The H$\alpha$ emission is smoothly distributed and somewhat extended, with the peak slightly offset from the optical center of the galaxy.

### 3.3 Field 3

The central galaxy of field 3, the S0a galaxy UGC 598, was not detected in H I. Two gas-rich galaxies were detected in this field, at large velocities relative to the central galaxy: $V_{\rm hel} = 5485\,{\rm km\,s^{-1}}$ and $5332\,{\rm km\,s^{-1}}$ for the H I-dwarfs UGC 598a and UGC 598b, respectively (see Fig. 1), compared to $5005\,{\rm km\,s^{-1}}$ for UGC 598 (obtained from RC3). The distribution of neutral hydrogen in UGC 598a is practically unresolved, and is somewhat offset from the optical centroid (Fig. 10). The H I in the other H I-dwarf UGC 598b seems to consist of two components, separated by about $2'$ but at the same redshift (Figs. 5 and 11). The smaller component of UGC 598b, though faint, is seen in three independent velocity channels and is not a noise artifact.

Both H I-dwarfs in this field have low central $B$-band surface brightnesses, UGC 598b being the faintest of all the H I-dwarfs in our sample in this respect. The optical morphology of UGC 598a is slightly asymmetrical with a roughly exponential surface brightness profile, and it has a smooth, centrally concentrated H$\alpha$ distribution. Seen at optical wavelengths, UGC 598b shows an irregular structure with no clearly defined center. The optical emission coincides with the peak of the larger H I component. The inclination, position angle, and radial color profiles of UGC 598b, derived by fitting ellipses to the optical images, are not very accurate due to the amorphous nature of this galaxy. The $B$-band image of UGC 598b could not be flat fielded properly; its total blue magnitude is therefore not well determined.

### 3.4 Field 4

The target galaxy of this field is UGC 752, an S0 galaxy. The only other 21-cm detection in this field is a galaxy from the Zwicky catalog, Zw 502.004 (Fig. 6). The central galaxy UGC 752 has an H I disk in differential rotation, as the double peaked-global H I profile in the center right panel of Figure 1 shows. The asymmetry of the line profile and the curving of the kinematical major axis (center left panel of Fig. 2) could indicate some disturbance in UGC 752. The H I distribution of Zw 502.004 is irregular in appearance and is slightly offset with respect to the optical image (Fig. 12). In spite of this, both the global profile (center right panel of Fig. 1) and the velocity field (center right panel of Fig. 2) are regular and symmetrical, with the iso-velocity contours of the gas aligned with the optical minor axis. The optical image of Zw 502.004 shows a normal disk galaxy, with an exponential light profile. No H$\alpha$ observations were made of the galaxies in this field.

## 3.5 Field 5

This field contains two optically bright galaxies, UGC 810 and UGC 820. The galaxies are classified as Sc and SBa-b, respectively. Despite their proximity to each other in position and velocity space, both galaxies have regular H I distributions (Fig. 7) and kinematics (bottom left panel of Fig. 1; bottom two panels of Fig. 2) and show no obvious signs of interaction. Solar interference during the VLA observations of this field made it necessary to discard a large fraction of the UV data. Consequently, the sensitivity in this field is lower than in other fields, with an r.m.s. noise level of 0.6 mJy/beam. We did not observe these two galaxies in H$\alpha$.

## 3.6 Fields 6, 7, and 8

H I-dwarfs were found in each of these three blank supercluster fields. Two of the detected H I-dwarfs, PPHI 0023+34 (field 6) and PPHI 0147+31 (field 8), have asymmetrical H I distributions that are offset slightly with respect to the optical image (Figs. 13 and 15). The northward extension of the gas in PPHI 0147+31 is seen in only one channel and may not be real. Both PPHI 0023+34 and PPHI 0147+31 are low surface brightness systems and have irregular optical morphology. The H I-dwarf detected in field 7, PPHI 0036+34, is somewhat more regular than the other two H I-dwarfs in both its H I and light distributions (Fig. 14). Its stellar disk has an exponential brightness profile. None of the three H I-dwarfs in the blank fields were observed in H$\alpha$.

## 4 Discussion

We discussed the spatial distribution of our H I-dwarfs at some length in Paper I; we have little to add here on the subject, since we have no new data on the spatial distribution. We found nine H I-dwarfs in the supercluster and none in the void, so the basic message is that H I-dwarfs trace the structure outlined by optically bright galaxies. The correlation seems to continue down to small scales within the supercluster, since we found six of our nine H I-dwarfs in the fields targeted on bright galaxies despite observing more blank fields than targeted fields. This is the sort of behavior we might expect based on the galaxy autocorrelation function, which continues as a power law down to very small scales. With only nine objects, we are unable to rule out a modest bias between the clustering of bright galaxies and the clustering of H I-dwarfs (see Paper I for details), but it is clear that the most likely place to find a gas-rich H I-dwarf is near an optically bright giant.

The optical and H$\alpha$ data reported in §3 allow us to compare the properties of our H I-selected sample to those of more traditional, optically selected samples. We provide such a comparison in §4.1 below. We then describe possible environmental effects within our sample and discuss the implications of our results for the H I mass density and cross-section in the local universe.

## 4.1 Comparison to Other Samples

### 4.1.1 Luminosities and Star Formation Rates

The distinguishing feature of our sample of H I-dwarfs is that it is not pre-selected in the optical, so the most interesting question to ask about it is whether the objects detected in this way are anomalously faint. There are at least two sensible ways to pose this question: are the H I-dwarfs unusually faint for their H I masses, and are they unusually faint for their circular velocities?

The first of these questions can be rephrased in terms of the H I-mass-to-light ratios $M_{\rm HI}/L_B$, which are listed in column 8 of Table 5. Values for the H I-dwarfs range from 0.08 to 3.23, with a mean of 1.17 (we exclude UGC 60a, for which we did not obtain $L_B$). For their sample of 29 irregular galaxies in the Virgo cluster (excluding five that were not detected in H I), Gallagher & Hunter (1989) find ratios ranging from 0.02 to 1.76, with a mean of 0.39. In Haynes & Giovanelli's (1984) sample of isolated galaxies, the $M_{\rm HI}/L_B$ ratios for the 46 galaxies of type later than Sc (specifically, those assigned a numerical type 8–10 in Table I of their paper) range from 0.15 to 9.8, with a mean of 1.34. Two galaxies in the Haynes & Giovanelli sample have $M_{\rm HI}/L_B$ larger than the extreme value in our sample, that of PPHI 0023+34. It appears that the $M_{\rm HI}/L_B$ ratios of our H I-selected galaxies are not extraordinarily high relative to those of optically selected samples of late-type galaxies. Certainly, 21-cm selection has not turned up any population resembling star-free gas clouds. There is some evidence for an increase in $M_{\rm HI}/L_B$ in lower density environments: our supercluster has a higher mean ratio than Gallagher & Hunter's Virgo cluster sample, Haynes & Giovanelli's sample of isolated galaxies has a still higher mean ratio, and within our sample, the blank-field H I-dwarfs have high $M_{\rm HI}/L_B$ relative to H I-dwarfs in target fields.

We can address the second of the questions posed above by plotting our sample of galaxies on the Tully-Fisher (1977) relation, as shown in Figure 16. The dashed line in this Figure represents the relation derived by Pierce & Tully (1988, hereafter PT). For consistency, we have corrected the absolute magnitudes $M_B^{b,i}$ listed in column 6 of Table 4 to the Hubble constant $H_0 = 85 \, {\rm km \, s^{-1} \, Mpc^{-1}}$ derived by PT. Linewidths, corrected for inclination and internal motions, are from column 10 of Table 3 (see §2.1). The H I-dwarfs in our sample follow the Tully-Fisher relation derived from optically selected galaxies fairly well; they are not unusually faint for their circular velocity. The three H I-dwarfs detected in the blank fields all lie somewhat below the PT relation (the possible implications of this are discussed in §4.2). An unweighted least-squares fit to all of the galaxies in our sample yields a slope of $-6.0 \pm 1.1$, more or less consistent with PT's value of $-7.30 \pm 0.32$. This fit is plotted as a dotted line in Figure 16. The vertical scatter about the best-fit relation is 1.0 magnitude. This scatter is higher than that found for optical samples, but it could arise largely from errors in the linewidths and inclinations of the small systems in our sample.

Kunth (1988) and Tyson & Scalo (1988) suggest that the majority of dwarf galaxies may be in a quiescent state, not forming young stars, and that they might consequently be missed by optical selection methods. We have already seen that the H I-dwarfs in our sample are not unusually faint, and we can address the star formation issue directly for the four Hı-dwarfs that we observed in Hα. The Hα equivalent widths (column 10 of Table 5) are similar to those of typical late-type galaxies (Kennicutt & Kent 1983), and the $\dot{M}/L_B$ ratios and the timescales for gas depletion (columns 12 and 13 of Table 5) are similar to the values derived by Gallagher & Hunter (1989) for their sample of Virgo irregulars. The gas depletion timescales are on the order of a Hubble time, so these Hı-dwarfs do not seem to be in the midst of star formation bursts.

In Paper I, we argued on the basis of the number of detected objects that our Hı-selection procedure had not turned up a population of optically under-luminous galaxies. The optical data presented here confirm this conclusion; our Hı-dwarfs appear to be garden variety late-type galaxies. If there is a large population of optically faint galaxies lurking in the shadows, they must either have very little gas, or they must store that gas in ionized or molecular form.

### 4.1.2 Colors and Ages

To our knowledge, no published $BRI$ magnitudes exist for large samples of dwarf galaxies. In Table 6, therefore, we compare the colors of our galaxies to those of the statistically complete, diameter-limited galaxy sample of de Jong and van der Kruit (in preparation). They have obtained $BVRIHK$ surface photometry of 86 galaxies, ranging in morphological type from Sa to Irr. Table 6 lists the mean and dispersion of the $B - R$ and $R - I$ colors for their sample (the "control sample") and for a subset of this sample with morphological types Sd and later. We also list the mean colors and dispersions in Table 6 for those galaxies in our sample which have $BRI$ photometry: eight of the nine Hı-dwarfs and three of the eight cataloged galaxies (see Table 4). Our sample is somewhat bluer than the control sample in $B - R$, but the difference is not large compared to the dispersion within the control sample. Hunter & Gallagher (1986) find that optically selected dwarf galaxies tend to be bluer than bright galaxies, and the same trend seems to hold for Hı-selected galaxies.

Figure 17 plots the galaxies for which we have $BRI$ photometry in a color-color diagram, with the Hı-dwarfs as open squares and the cataloged galaxies as filled triangles. For comparison, the open circles and connecting line show the galaxy evolution model of Mazzei et al. (1992). The model incorporates stellar evolution, chemical evolution of the stellar population, and absorption and re-emission of starlight by interstellar dust. The model does not fit the observational data very well, and two colors are in any case not enough for accurate age determination. However, it does appear that the galaxies in our sample are fairly young, and that most of the Hı-selected galaxies are younger than the cataloged galaxies.

### 4.2 Environmental Effects

In Figure 18, we plot the central $B$ surface brightness (upper panel) and the logarithm of the total blue luminosity (lower panel) of all galaxies that were not field targets against the logarithm of the projected distance to the field's target galaxy (or, in the case of the blank-field Hı-dwarfs, the projected distance to the nearest galaxy in HG's redshift catalog). Cataloged galaxies are shown by triangles, Hı-dwarfs by squares. Surface brightnesses come from column 3 of Table 5, and luminosities from column 11 of Table 4 (see §2.2 for details). In both cases, there appears to be a modest trend, with fainter galaxies having larger projected separations.

We can make this trend a bit more convincing if we distinguish between HI-dwarfs that are gravitationally bound to bright galaxies and HI-dwarfs that are not. We can reasonably assume that the HI-dwarfs in blank fields are not bound, but the question is trickier for HI-dwarfs in target fields. We will try to make the distinction by computing dynamical mass-to-light ratios on the assumption that the HI-dwarfs are bound, then asking whether the resulting values are "plausible."

We use two different techniques to estimate mass-to-light ratios. The first employs the "average" mass estimator of Heisler et al. (1985) for galaxy groups:

$$M_{\text{est}} = \frac{2 f_{\text{Av}}}{GN(N-1)} \sum_i \sum_{j<i} (V_{zi} - V_{zj})^2 R_{\perp,ij} \quad .$$

Here $N$ is the number of galaxies in the group, $V_{zi}$ is the velocity along the line of sight, $R_{\perp,ij}$ is the projected distance between galaxies $i$ and $j$, $G$ is the gravitational constant, $f_{\text{Av}}$ is a scaling factor, and the sums are carried out over all group members. Heisler et al. (1985) calibrate this and other, similar mass estimators using model groups with $N = 5$ and $N = 10$. For $N = 5$, they find that 75% of the mass estimates lie within a factor $10^{0.25}$ of the correct value. Applying the estimator to the nearby groups cataloged by Huchra & Geller (1982), they find a median $M/L$ of about $10^{2.8}\, M_\odot/L_\odot$ for groups with three members.

The mass estimates for our five groups (*i.e.* our five target fields) appear in column 5 of Table 7. The corresponding mass-to-light ratios appear in column 7. While four of the five groups have what might be called reasonable $M/L$ values, the high value in field 3, 2931 $M_\odot/L_\odot$, makes it unlikely that the galaxies in this field form a gravitationally bound group. Even for the bound groups, the mass estimates in Table 7 have large statistical uncertainties because of the small numbers of group members.

To avoid these large uncertainties, we can take a more conservative approach and compute the minimum mass that would be required to bind each companion to its bright target galaxy. We obtain this mass by assuming that the line-of-sight velocity difference $\Delta V$ and projected radial distance $\Delta R$ represent the true, three-dimensional values. These quantities are listed in columns 3 and 4 of Table 7. The minimum mass follows from the inequalities

$$KE \leq -PE \quad \Longrightarrow \quad \frac{M_1 M_2}{2(M_1 + M_2)}(\Delta V)^2 \leq \frac{GM_1 M_2}{\Delta R} \quad ,$$

where $M_1$ and $M_2$ are the masses of the central galaxy and the companion, respectively. This inequality implies a minimum value of $(M_1 + M_2)$:

$$M_{\min} = \frac{1}{2G}(\Delta V)^2 \Delta R \quad .$$

Column 6 of Table 7 lists this minimum pair mass for each companion galaxy (plus the central galaxy), and column 8 lists the mass-to-light ratio obtained by dividing this minimum mass by the total blue luminosity of the pair. Again, we conclude that the galaxies in field 3 are probably not gravitationally bound, while the remaining galaxies could plausibly be bound.

This conclusion has an interesting implication for the low surface brightness systems in our sample. Three of the HI-dwarfs have very low surface brightness ($\mu_0 > 23$); two of these were found in blank fields, the third in field 3. This point is illustrated in Figure 18, where we use open symbols to denote "unbound" systems, *i.e.* blank-field HI-dwarfs and the HI-dwarfs in field 3, and filled symbols to denote bound systems. The average $\mu_0$ of the unbound systems is 22.7, with a dispersion of 1.3; for the bound systems, the average is 20.6, with a dispersion of 0.7. The mean and dispersion in values of $\log(L_B)$ are $8.61 \pm 0.37$ for the unbound systems and $9.43 \pm 0.39$ for the bound systems.

Although the statistics are poor, there is at least a hint in our data that the presence of a neighboring bright galaxy influences surface brightness and luminosity. In particular, all of our low surface brightness systems are unbound, consistent with the findings of Bothun et al. (1993), though not all of our isolated systems have low surface brightness.

## 4.3 HI Cross Section and Mean Density in the Local Universe

Our 21-cm survey provides us with a picture of the distribution of neutral gas in regions of the nearby universe. This picture can be compared to the distribution of neutral hydrogen inferred from quasar absorption lines at high redshifts. Most relevant to our survey are the damped Ly$\alpha$ absorbers (Wolfe 1990; Lanzetta et al. 1991), which have neutral hydrogen column densities similar to those of the systems that we detect at 21 cm. The question of what constitutes the parent population of the damped Ly$\alpha$ systems remains a matter of debate. It has been suggested that the absorption lines could arise in extended galactic disks (Wolfe 1990), galactic halos (Bahcall & Spitzer 1969), dwarf galaxies (York et al. 1986; Tyson 1988), or tidal features around interacting and merging galaxies (Carilli & van Gorkom 1992).

By most accounts (*cf.* Lanzetta et al. 1991; Rao & Briggs 1993), the neutral hydrogen contained in damped Ly$\alpha$ systems greatly exceeds that in present-day galaxies, by factors of $\sim$ 5–10. However, the estimate of the hydrogen content at the current epoch is based on observations of gas associated with optically cataloged galaxies. By comparing theoretical models to observations, Tyson & Scalo (1988) conclude that many dwarf galaxies might escape optical detection. Tyson (1988) argues that this population of faint dwarfs could produce the observed damped Ly$\alpha$ absorption *without* strong evolution in the neutral gas fraction.

Arguments like those of Tyson & Scalo (1988) provided much of the motivation for our 21-cm survey. However, we found in Paper I that the Tyson & Scalo (1988) model predicted far more dwarfs than we observed, unless we assumed that the ratio of HI mass to optical luminosity was much *smaller* for "quiescent" dwarfs than for bright HI-dwarfs. In a similar vein, we have seen in §4.1 that our HI-selected galaxies are not anomalously faint, so estimates of the local neutral gas density based on optically selected samples, *à la* Lanzetta et al. (1991) and Rao & Briggs (1993), should be quite safe. Our survey is not sensitive to HI masses much below $10^8 M_\odot$, but other "blind" 21-cm surveys have covered this regime. Although these surveys cover smaller volumes than our survey, they are probably large enough to rule out a steep rise in the H I mass function between $10^8 \, M_\odot$ and $\sim 10^6 \, M_\odot$ (Fisher & Tully 1981; Hoffman et al. 1989; Briggs 1990; Hoffman et al. 1992).

We could attempt to estimate the space density and cross-section of neutral hydrogen in the local universe directly from our sample. Unfortunately, our survey volume is small, and our fields are not randomly chosen, so such an estimate would have large statistical uncertainty and a systematic bias that would be difficult to evaluate. However, we can say that within our survey region, the H I-dwarfs contribute only 19% of the H I mass and 12% of the H I cross-section. Once again, the message is that the H I content of the nearby universe is dominated by relatively bright galaxies. Lanzetta (1993) notes the suggestive coincidence that the mass seen in gas at high redshifts is equal to the luminous mass seen at $z = 0$, and he therefore proposes that the gas observed in the damped Ly$\alpha$ systems has formed into the stars of present-day galaxies. Our results support this inference, albeit indirectly.

### 4.4 Summary

Galaxies can "shine" at both optical and 21-cm wavelengths, and in principle one might find rather different objects in these two regions of the spectrum. However, our H I survey of the Perseus-Pisces supercluster and void has not turned up any remarkable population of unusual objects. Most of the neutral hydrogen gas is associated with big, optically bright galaxies. Sensitive 21-cm observations reveal a substantial number of smaller objects, but these appear to be normal irregular galaxies, and their spatial distribution is not radically different from that of the optically bright giants. The distribution of H I in an individual galaxy or galaxy group may be quite different from the distribution of stars, but in its broad features, the universe seen at 21 cm is a familiar one.

*Acknowledgments.* We are grateful to Martha Haynes and Riccardo Giovanelli for helpful discussions and for allowing us to use their redshift data, in advance of publication, to select our fields. We also thank Richard McMahon for providing us with APM data that were used to determine star positions on some of our JKT frames, and Ronald Hes for excellent support during the early phase of this project. We thank Hy Spinrad and Dave Cudaback for the loan of narrow-band filters used to make the H$\alpha$ measurements. Finally, we thank the NRAO for generous allocation of observing time. This research was supported in part by NSF grants AST 89-17744 and AST 90-23254 to Columbia University, under grant no. 782-373-046 by the Netherlands Foundation for Research in Astronomy (NFRA), which receives its funds from the Netherlands Organization for Scientific Research (NWO). ASF and PG were supported by NASA through grants HF-1005.01-90A and HF-1033.01-92A, respectively, from the Space Telescope Science Institute, which is operated by the Association of Universities for Research in Astronomy, Inc., under NASA contract NAS5-26555. DW acknowledges support from the W.M. Keck Foundation and from NSF grant PHY 92-45317. The Jacobus Kapteyn Telescope (JKT) and the William Herschel Telescope (WHT) are operated on the island of La Palma by the Royal Greenwich Observatory in the Spanish Observatorio del Roque de los Muchachos of the Instituto de Astrofísica de Canarias.

# References

Bahcall, J.N., and Spitzer, L. Jr. 1969, ApJ, 156, L63.

Bothun, G.D., Schombert, J.M., Impey, C.D., Sprayberry, D., and McGaugh, S.S. 1993, AJ, 106, 530.

Bottinelli, L., Gouguenheim, L., Fouqué, P., Paturel, G. 1990, A&AS, 82, 391.

Briggs, F.H. 1990, AJ, 100, 999.

Burstein, D., and Heiles, C. 1984, ApJS, 54, 33.

Carilli, C.L., and van Gorkom, J.H. 1992, ApJ, 399, 373.

Fisher, J.R., and Tully, R.B. 1981, ApJ, 243, L23.

Gallagher, J.S. and Hunter, D.A. 1989, AJ 98, 806.

Haynes, M.P., and Giovanelli, R. 1984, AJ, 89, 758.

Haynes, M.P., and Giovanelli, R. 1989, in Pont. Acad. Sci. Study Week 27, Large-Scale Motions in the Universe, ed. V. Rubin and G. Coyne (Rome: Specola Vaticana), 31 (HG).

Heisler, J., Tremaine, S., and Bahcall, J.N. 1985, ApJ, 298, 8.

Henning, P.A. 1992, ApJS, 78, 365.

Hoffman, G.L, Helou, G., Salpeter, E.E., and Williams, H.L. 1989, ApJS, 69, 65.

Hoffman, G.L., Lu, N.Y., and Salpeter, E.E. 1992, AJ, 104, 2086.

Huchra, J.P., and Geller, M.J. 1982, ApJ, 257, 423.

Hunter, D.A., and Gallagher, J.S. 1986, PASP, 98, 5.

Kennicutt, R.C., and Kent, S.M. 1983, AJ, 88, 1094.

Kunth, D. 1988, in Evolutionary Phenomena in Galaxies, ed. J.E. Beckman and B.E.J. Pagel (Cambridge: Cambridge University Press), 22.

Landolt, A.U. 1983, AJ, 88, 439.

Lanzetta, K.M., Wolfe, A.M., Turnshek, D.A., Lu, L., McMahon, R.G., and Hazard, C. 1991, ApJS, 77, 1.

Lanzetta, K.M. 1993, in The Evolution of Galaxies and Their Environment, Proceedings of the Third Tetons Summer Astrophysics Conference, ed. J.M. Shull and H.A. Thronson, Jr. (Dordrecht: Kluwer Academic Publishers), 237.

Mazzei, P., Xu, C., and De Zotti, G. 1992, A&A, 256, 45.

Napier, P.J., Thompson, A.R., and Ekers, R.D. 1983, Proc. Inst. Electr. Eng., 71, 1295.

Neugebauer, G. et al. 1984, ApJ, 278, L1.

Nilson, P. 1973, Uppsala General Catalogue of Galaxies, Uppsala: Uppsala Obs. Ann. (UGC).

Pierce, M.J., and Tully, R.B. 1988, ApJ, 330, 579 (PT).

Rao, S., and Briggs, F.H. 1993, ApJ, 419, in press.

Rieke, G.H., and Lebofsky, M.J. 1985, ApJ, 288, 618.

Salpeter, E.E. 1955, ApJ, 121, 161.

Tyson, N.D. 1988, ApJ, 329, L57.

Tyson, N.D., and Scalo, J.M. 1988, ApJ, 329, 618.

Tully, B. 1988, Nearby Galaxies Catalog, Cambridge University Press.

Tully, R.B., and Fisher, J.R. 1977, A&A, 54, 661.

Tully, B., and Fouqué, P. 1985, ApJS, 58, 67.

de Vaucouleurs, G., de Vaucouleurs, A., Corwin, H.G. Jr., Buta, R.J., Paturel, G., and Fouqué, P. 1991, Third Reference Catalogue of Bright Galaxies, New York: Springer-Verlag (RC3).

Weinberg, D.H., Szomoru, A., Guhathakurta, P., and van Gorkom, J.H. 1991, ApJ, 372, L13 (Paper I).

Wolfe, A.M. 1990, in The Interstellar Medium in Galaxies, Proceedings of the Second Tetons Summer Astrophysics Conference, ed. H.A. Thronson, Jr. and J.M. Shull (Dordrecht: Kluwer Academic Publishers), 387.

York, D.G., Dopita, M., Green, R., and Bechtold, J. 1986, ApJ, 311, 610.

Zwicky, F., Herzog, E., Wild, P., Karpowicz, M., and Kowal, C.T. 1961–1968, Catalog of Galaxies and Clusters of Galaxies, Pasadena: California Inst. of Technology.

**Figure captions**

**Figure 1.** Global H I profiles of all detected galaxies. The flux densities have been corrected for primary beam attenuation. All galaxies detected in a given target field are plotted together in the same panel (fields 1–5 appear in order starting from top left), as are the three H I-dwarfs detected in blank fields (bottom right panel). The velocity resolution is about 43 km s$^{-1}$. The uncertainties in the flux densities per channel range from 0.4 to 2.5 mJy for the H I-dwarfs, depending on the distance to the field center, and from 0.5 to 1.8 mJy for the bright galaxies.

**Figure 2.** Intensity-weighted mean velocity fields of selected galaxies, those in which the H I is well resolved, both spatially and kinematically. These include the target galaxies of fields 1, 2, 4, and 5 (UGC 598, the target galaxy of field 3, was not detected in H I), along with Zw 502.004 (field 4) and UGC 810, the second bright galaxy in field 5. The spacing between adjacent iso-velocity contours is 20 km s$^{-1}$, except in the cases of UGC 60 and UGC 820 which have a spacing of 40 km s$^{-1}$, and the velocities of a few contours are marked in km s$^{-1}$. The FWHM of the synthesized VLA beam is indicated by the shaded ellipse in the lower left corner of each panel. In top right panel, the H I-dwarf UGC 60a can be seen as a close companion, 2′ south of the target galaxy UGC 60. The velocity field shows two distinct rotating disks whose projected rotation axes are perpendicular.

**Figure 3.** Spatial distribution of the total H I emission in field 1, centered on the target galaxy UGC 26. The map has not been corrected for primary beam attenuation. The FWHM of the beam is indicated by the shaded ellipse. The contour levels are 85 (equal to 3$\sigma$ in the outer parts of UGC 26), 170, 256, 426, 852, and 1704 mJy beam$^{-1}$ km s$^{-1}$.

**Figure 4.** Same as Figure 3 for field 2, centered on UGC 60. Note that the H I distributions of UGC 60 and the H I-dwarf UGC 60a overlap. The contour levels are 77 (equal to 2$\sigma$ in the outer part of UGC 60), 153, 230, 307, 511, 1022, 1534, 2045, 2556, and 3067 mJy beam$^{-1}$ km s$^{-1}$.

**Figure 5.** Same as Figure 3 for field 3, centered on the target galaxy UGC 598. No H I is detected in the target galaxy of this field, UGC 598; its position is marked by a cross. The contour levels are 47 (equal to 1.5$\sigma$ in the outer parts of UGC 598a), 94, 141, 187, and 234 mJy beam$^{-1}$ km s$^{-1}$.

**Figure 6.** Same as Figure 3 for field 4, centered on the target galaxy UGC 752. The contour levels are 47 (equal to 1.5$\sigma$ in the outer parts of UGC 752), 94, 141, 187, 234, 320, 426, 533, and 639 mJy beam$^{-1}$ km s$^{-1}$.

**Figure 7.** Same as Figure 3 for field 5, centered on the target galaxy UGC 820. The contour levels are 145 (equal to 2.2$\sigma$ in the outer parts of UGC 820), 290, 435, 963, 1444,

1926, 2407, and 2888 mJy beam$^{-1}$ km s$^{-1}$.

**Figure 8.** Contour plots of the total H I distribution of the galaxies in field 1 superimposed on $B$-band (negative) gray-scale images. The H I contours have been corrected for primary beam attenuation. The beam size (FWHM) is indicated by the shaded ellipse in each panel. The lowest gray level corresponds to $B \sim 25$ mag arcsec$^{-2}$. The H I contour levels are:
  (1) Zw 499.039: 3.7 (1.1$\sigma$), 7.4, 11.1, 14.8, and 18.5 $\times$ 10$^{19}$ cm$^{-2}$;
  (2) UGC 26a: 2.1 (1.8$\sigma$), 4.2, 6.3, 8.4, 10.5, and 12.6 $\times$ 10$^{19}$ cm$^{-2}$;
  (3) UGC 26: 1.2 (1.4$\sigma$), 3.7, 6.1, 8.5, 11, 18.3, 24.4 and 30.5 $\times$10$^{19}$ cm$^{-2}$; and
  (4) UGC 26b: 2.57 (2$\sigma$), 5.1, 7.7, 10.3, 12.9, 18.4, and 23.5 $\times$ 10$^{19}$ cm$^{-2}$.

**Figure 9.** Same as Figure 8 for UGC 60b, an H I-dwarf detected in field 2. The H I column density contours are 1.5 (1.5$\sigma$), 3, 4.5, 6 and 7.5 $\times$ 10$^{19}$ cm$^{-2}$.

**Figure 10.** Same as Figure 8 for UGC 598a, an H I-dwarf detected in field 3. The H I column density contours are 4.2 (1.6$\sigma$), 8.4, 12.6, and 16.8 $\times$ 10$^{19}$ cm$^{-2}$.

**Figure 11.** Same as Figure 8 for UGC 598b, an H I-dwarf detected in field 3. The H I column density contours are 1.5 (1.7$\sigma$), 3, 4.5, 6, 7.5, and 9 $\times$ 10$^{19}$ cm$^{-2}$.

**Figure 12.** Same as Figure 8 for Zw 502.004, a previously cataloged galaxy that was detected in H I in field 4. The H I column density contours are 1.4 (1.7$\sigma$), 2.8, 4.2, 5.6, 7, 9.6, and 12.8 $\times$ 10$^{19}$ cm$^{-2}$.

**Figure 13.** Same as Figure 8 for PPHI 0023+34, an H I-dwarf detected in blank field 6. The H I column density contours are 1.7 (1.8$\sigma$), 3.4, 5.1, 6.8, 8.5, 10.2, and 11.9$\times$10$^{19}$ cm$^{-2}$.

**Figure 14.** Same as Figure 8 for PPHI 0036+34, an H I-dwarf detected in blank field 7. The H I column density contours are 2.9 (1.6$\sigma$), 5.8, 8.7, 11.6, 14.5, 17.4, 20.3, and 23.2 $\times$ 10$^{19}$ cm$^{-2}$.

**Figure 15.** Same as Figure 8 for PPHI 0147+31, an H I-dwarf detected in blank field 8. The H I column density contours are 1.7 (1.5$\sigma$), 3.4, 5.1, 6.8, and 8.5 $\times$ 10$^{19}$ cm$^{-2}$.

**Figure 16.** The $B$-magnitude-vs-linewidth (Tully-Fisher) relation. The dashed line shows the relation derived for optically selected galaxies by Pierce and Tully (1988), the dotted line an unweighted least-squares fit to our data. Triangles represent the previously cataloged galaxies in our H I sample, squares the H I-dwarfs. Absolute magnitudes are computed assuming $H_0 = 85$ km s$^{-1}$ Mpc$^{-1}$, for consistency with Pierce and Tully. Our H I-dwarfs lie close to the relation derived for optically selected galaxies, indicating that they are not anomalously faint for their circular velocities.

**Figure 17.** Two-color diagram for galaxies in our sample. Squares represent HI-dwarfs; triangles represent the three previously cataloged galaxies for which we have $BRI$ photometry. The open circles and connecting line show the evolution model of Mazzei et al. (1992).

**Figure 18.** Extrapolated central $B$-band surface brightness (upper panel) and logarithm of total blue luminosity (lower panel) plotted against logarithm of the projected angular distance to the nearest bright galaxy. Triangles represent previously cataloged galaxies (excluding field target galaxies); we have a blue luminosity but not a central surface brightness for UGC 810. Squares represent HI-dwarfs, which we have divided into bound (filled symbols) and unbound (open symbols) systems as described in §4.3. There is weak evidence for an environmental effect—in particular, the three low surface brightness systems are all unbound.

Table 1: Parameters of VLA observations.

| | | |
|---|---|---|
| Configuration | | D |
| Date of observations | | November 1989 |
| Number of antennae | | 27 |
| Observing mode | | 2IF |
| Total bandwidth | | 6.25 MHz |
| Number of channels | | 31 |
| Shortest spacing | | 0.033 km |
| Longest spacing | | 1.03 km |
| FWHM of synthesized beam | | $\sim 70'' \times 60''$ |
| FWHM of primary beam | | $30'$ |
| Equivalent $T_b$ for 1 mJy/beam | | 0.15 K |
| Region | Void | Supercluster |
| Central heliocentric velocity | 3250 km s$^{-1}$ | 5100 km s$^{-1}$ |
| Velocity resolution | 42.1 km s$^{-1}$ | 42.6 km s$^{-1}$ |
| Velocity coverage | 1220.9 km s$^{-1}$ | 1235.4 km s$^{-1}$ |
| Number of fields | 30 | 12 |
| Integration time per field | 40 min | 210 min |
| R.m.s. noise | 1 mJy/beam | 0.4 mJy/beam |

Table 2: Pointings of VLA observations.

| (1) $\alpha_{1950}$ (h m s) | (2) $\delta_{1950}$ ( ° ′ ″) | (3) Region |
|---|---|---|
| 0 00 00.0 | 32 00 00 | V+SC |
| 0 01 48.0 | 31 12 00 | V+SC |
| 0 04 48.0 | 32 19 48 | V+SC |
| 0 12 00.0 | 32 00 00 | V+SC |
| 0 24 00.0 | 30 00 00 | V |
| 0 24 00.0 | 32 00 00 | V |
| 0 24 00.0 | 34 00 00 | V+SC |
| 0 36 00.0 | 30 00 00 | V |
| 0 36 00.0 | 32 00 00 | V |
| 0 36 00.0 | 34 00 00 | V+SC |
| 0 48 00.0 | 30 00 00 | V |
| 0 48 00.0 | 32 00 00 | V |
| 0 48 00.0 | 34 00 00 | V |
| 0 55 12.0 | 31 12 36 | SC |
| 0 60 00.0 | 30 00 00 | V |
| 0 60 00.0 | 32 00 00 | V |
| 0 60 00.0 | 34 00 00 | V |
| 1 09 36.0 | 31 51 36 | V+SC |
| 1 12 00.0 | 30 00 00 | V |
| 1 12 00.0 | 32 00 00 | V |
| 1 12 00.0 | 34 00 00 | V |
| 1 13 12.0 | 30 46 12 | V+SC |
| 1 24 00.0 | 30 00 00 | V+SC |
| 1 24 00.0 | 32 00 00 | V |
| 1 24 00.0 | 34 00 00 | V |
| 1 28 48.0 | 33 21 36 | V |
| 1 36 00.0 | 30 00 00 | V+SC |
| 1 36 00.0 | 32 00 00 | V |
| 1 36 00.0 | 34 00 00 | V |
| 1 48 00.0 | 32 00 00 | V+SC |
| 1 57 36.0 | 31 11 24 | V |

Note to TABLE 2:
Col. 3: V denotes the void fields, SC the supercluster fields.

**Table 3.** HI and radio continuum parameters

| (1) Field | (2) Name | (3) Type | (4) $\alpha_{1950}$ (h m s) | (5) $\delta_{1950}$ (° ′ ″) | (6) $V_{\rm hel}$ (km/s) | (7) $D$ (Mpc) | (8) $W_{20}$ (km/s) | (9) $W_{50}$ (km/s) | (10) $\log W_{\rm R}^{\rm i}$ | (11) $M_{\rm HI}$ ($10^8 M_\odot$) | (12) $F_{\rm cont}$ (mJy) |
|---|---|---|---|---|---|---|---|---|---|---|---|
| 1 | Zw 499.039 | Sc | 0 00 15.4 | 31 12 13 | $4952 \pm 10$ | 51.94 | $109 \pm 20$ | $77 \pm 20$ | 2.20 | $7.1 \pm 1.8$ | |
| | UGC 26a | | 0 01 13.2 | 31 03 54 | $4997 \pm 10$ | 52.38 | $173 \pm 20$ | $118 \pm 20$ | 2.35 | $4.5 \pm 0.8$ | |
| | UGC 26* | SBb | 0 01 50.2 | 31 11 39 | $4954 \pm 10$ | 51.96 | $289 \pm 20$ | $259 \pm 20$ | 2.45 | $65.9 \pm 4.6$ | $7.7 \pm 0.6$ |
| | UGC 26b | | 0 02 00.8 | 31 26 01 | $4845 \pm 10$ | 50.87 | $149 \pm 20$ | $123 \pm 20$ | 2.16 | $7.8 \pm 1.5$ | |
| 2 | UGC 60* | Sb | 0 04 44.7 | 32 19 49 | $5076 \pm 10$ | 53.17 | $524 \pm 20$ | $483 \pm 20$ | 2.67 | $37.2 \pm 1.6$ | $12.4 \pm 0.4$ |
| | UGC 60a | | 0 04 44.8 | 32 17 50 | $5005 \pm 12$ | 52.47 | $165 \pm 46$ | $94 \pm 20$ | | $4.1 \pm 1.2$ | |
| | UGC 60b | | 0 05 31.2 | 32 21 15 | $5124 \pm 10$ | 53.66 | $153 \pm 20$ | $101 \pm 20$ | 2.03 | $2.2 \pm 0.4$ | |
| 3 | UGC 598a | | 0 54 02.4 | 30 58 29 | $5485 \pm 12$ | 56.97 | $121 \pm 41$ | $66 \pm 28$ | 1.93 | $7.0 \pm 3.0$ | |
| | UGC 598* | S0a | 0 55 06.3 | 31 12 53 | 5005 | 52.16 | | | | | |
| | UGC 598b | | 0 55 55.8 | 31 05 40 | $5332 \pm 10$ | 55.42 | $109 \pm 20$ | $80 \pm 20$ | 1.84 | $4.6 \pm 1.3$ | |
| 4 | UGC 752* | S0 | 1 09 22.4 | 31 51 28 | $5010 \pm 10$ | 52.13 | $351 \pm 20$ | $289 \pm 34$ | | $6.1 \pm 0.7$ | |
| | Zw 502.004 | | 1 09 23.7 | 31 44 17 | $5227 \pm 10$ | 54.30 | $239 \pm 20$ | $215 \pm 20$ | 2.35 | $7.3 \pm 0.9$ | |
| 5 | UGC 810 | Sc | 1 13 03.1 | 30 49 00 | $4831 \pm 10$ | 50.30 | $338 \pm 20$ | $273 \pm 20$ | 2.45 | $30.9 \pm 2.2$ | $5.1 \pm 0.9$ |
| | UGC 820* | SBa-b | 1 13 27.3 | 30 46 00 | $4970 \pm 10$ | 51.68 | $505 \pm 21$ | $470 \pm 20$ | 2.66 | $48.5 \pm 2.6$ | $36.1 \pm 0.9$ |
| 6 | PPHI 0023+34 | | 0 23 23.8 | 34 11 32 | $4865 \pm 10$ | 51.01 | $151 \pm 20$ | $84 \pm 22$ | 2.05 | $6.1 \pm 1.0$ | |
| 7 | PPHI 0036+34 | | 0 36 48.6 | 34 14 24 | $4796 \pm 12$ | 50.24 | $214 \pm 28$ | $164 \pm 37$ | 2.39 | $10.3 \pm 1.9$ | |
| 8 | PPHI 0147+31 | | 1 47 13.6 | 31 49 09 | $5039 \pm 10$ | 52.17 | $117 \pm 24$ | $67 \pm 24$ | 1.94 | $2.9 \pm 0.8$ | |

Notes to TABLE 3:

Col. 2: Target galaxies are denoted by asterisks. UGC 598, the target galaxy of field 3, was not detected in H I: all data related to this galaxy and all optical data related to the cataloged galaxies which were not reobserved by us were taken from the Third Reference Catalogue of Bright Galaxies (RC3; de Vaucouleurs et al. 1991).

Col. 6: Heliocentric velocity, computed by averaging the mean velocities at 20% and 50% of the peak emission. The uncertainties are derived from the uncertainties in the 21 cm fluxes per channel, but estimated to be no less than $10\,{\rm km\,s^{-1}}$.

Col. 7: Adopted distance, calculated from the heliocentric velocity corrected for solar motion with respect to the velocity centroid of the Local Group.

Col. 8 and 9: 21 cm line widths at 20% and 50% of the peak emission, not corrected for instrumental broadening. The uncertainties are derived from the uncertainties in the 21 cm fluxes per channel, but assumed to be no less than $20\,{\rm km\,s^{-1}}$.

Col. 10: Logarithm of the 21 cm line width at 20% of the peak emission, corrected for instrumental broadening according to Bottinelli et al. (1990), for internal motions according to Tully & Fouqué (1985) and corrected to edge-on. We do not list this parameter for UGC 752 as it is face-on.

Col. 11: Total H I mass. Errors are derived from the $1\sigma$ errors in the 21 cm flux per channel and the uncertainties in velocity width and distance.

Col. 12: Integrated 21 cm continuum flux at the position of the H I emission. The errors are the $1\sigma$ errors in the continuum maps.

**Table 4.** Optical parameters

| (1) Field | (2) Name | (3) $B_T$ (mag) | (4) $B-R$ (mag) | (5) $R-I$ (mag) | (6) $M_B^{b,i}$ (mag) | (7) $(B-R)^{b,i}$ (mag) | (8) $(R-I)^{b,i}$ (mag) | (9) $i$ (°) | (10) P.A. (°) | (11) $L_B$ ($10^8 L_{B,\odot}$) |
|---|---|---|---|---|---|---|---|---|---|---|
| 1 | Zw 499.039 | 15.44 ± 0.31 | 1.23 ± 0.36 | 0.56 ± 0.21 | −18.3 | 1.15 | 0.52 | 25.0 ± 5.0 | 14 ± 16 | 33.6 ± 9.7 |
|   | UGC 26a | 16.88 ± 0.05 | 0.68 ± 0.07 | 0.72 ± 0.07 | −16.9 | 0.59 | 0.68 | 31.8 ± 2.5 | 119 ± 10 | 9.2 ± 0.5 |
|   | UGC 26* | 14.18 ± 0.05 | 0.97 ± 0.07 | 0.68 ± 0.07 | −19.7 | 0.83 | 0.61 | 54.2 ± 2.4 | 101 ± 5 | 122 ± 6 |
|   | UGC 26b | 16.47 ± 0.05 | 0.94 ± 0.07 | 0.73 ± 0.07 | −17.3 | 0.83 | 0.68 | 43.3 ± 1.8 | 147 ± 2 | 13.3 ± 0.6 |
| 2 | UGC 60* | 13.60 ± 0.16 |   |   | −20.8 |   |   | 78.8 | 38 | 321 ± 48 |
|   | UGC 60a |   |   |   |   |   |   |   |   |   |
|   | UGC 60b | 16.10 ± 0.13 | 1.24 ± 0.19 | 0.34 ± 0.17 | −18.2 | 0.99 | 0.23 | 73.5 ± 0.3 | 76 ± 1 | 28.7 ± 3.5 |
| 3 | UGC 598a | 17.43 ± 0.06 | 0.98 ± 0.09 | 0.56 ± 0.08 | −16.8 | 0.78 | 0.47 | 65.9 ± 1.7 | 103 ± 1 | 8.4 ± 0.5 |
|   | UGC 598* | 14.56 ± 0.15 |   |   | −19.7 |   |   | 73.8 | 29 | 117 ± 16 |
|   | UGC 598b | 18.76 ± 0.23 | 1.09 ± 0.25 | 0.30 ± 0.10 | −15.7 | 0.80 | 0.18 | 74.9 | 65 | 2.9 ± 0.6 |
| 4 | UGC 752* | 13.09 ± 0.11 |   |   | −20.7 |   |   | 3 | 0 | 286 ± 29 |
|   | Zw 502.004 | 15.65 ± 0.05 | 1.29 ± 0.07 | 0.54 ± 0.07 | −18.3 | 1.16 | 0.48 | 53.3 ± 1.1 | 7 ± 1 | 33.8 ± 1.6 |
| 5 | UGC 810 | 14.53 ± 0.12 |   |   | −19.8 |   |   | 79.8 | 157 | 133 ± 15 |
|   | UGC 820* | 13.76 ± 0.13 |   |   | −20.6 |   |   | 77.9 | 43 | 267 ± 32 |
| 6 | PPHI 0023+34 | 18.81 ± 0.06 | 1.03 ± 0.09 | 0.48 ± 0.08 | −15.2 | 0.83 | 0.39 | 63.5 ± 1.4 | 72 ± 2 | 1.9 ± 0.1 |
| 7 | PPHI 0036+34 | 16.38 ± 0.05 | 1.28 ± 0.08 | 0.64 ± 0.08 | −17.4 | 1.16 | 0.59 | 39.7 ± 2.4 | 45 ± 4 | 14.2 ± 0.7 |
| 8 | PPHI 0147+31 | 18.80 ± 0.06 | 0.95 ± 0.08 | 1.01 ± 0.08 | −15.1 | 0.83 | 0.96 | 58.1 ± 2.9 | 68 ± 6 | 1.7 ± 0.1 |

Notes to TABLE 4:
Col. 2: Target galaxies are denoted by asterisks. UGC 598, the target galaxy of field 3, was not detected in H I: all data related to this galaxy and all optical data related to the cataloged galaxies which were not reobserved by us were taken from the RC3 catalog.
Col. 3: Total $B$-band magnitude. Errors are a combination of the $1\sigma$ zero-point errors, flat-fielding errors and the estimated errors due to imperfect subtraction of adjacent and superposed stars.
Col. 4 and 5: Uncorrected broadband colors. Errors are derived from the errors in the total magnitudes in the individual bands.
Col. 6: Absolute $B$ magnitude corrected for Galactic and internal absorption following Burstein & Heiles (1984) and Tully & Fouqué (1985).
Col. 7 and 8: Broadband colors corrected for Galactic and internal absorption following Burstein & Heiles (1984) and Tully & Fouqué (1985).
Col. 9 and 10: Inclinations and position angles derived by fitting ellipses to the $I$-band images, assuming an intrinsic axis ratio of 0.2 and correcting for the 3° effect (Tully 1988). These quantities are not well determined for Zw 499.039 and UGC 598b. The values for UGC 60, UGC 598, UGC 752, UGC 810 and UGC 820 are taken from the RC3 catalog. The errors are the r.m.s. variation of the fits.
Col. 11: Total blue luminosity, derived from the blue magnitudes listed in column 6, assuming an absolute solar blue magnitude of 5.48.

**Table 5.** H$\alpha$ and derived parameters

| (1) Field | (2) Name | (3) $\mu_{0,B}$ $\left(\frac{\text{mag}}{\text{arcsec}^2}\right)$ | (4) $\mu_{0,B}^{b,i}$ $\left(\frac{\text{mag}}{\text{arcsec}^2}\right)$ | (5) $h$ ($''$) | (6) $D_{25}$ ($''$) | (7) $D_{\text{HI}}$ ($''$) | (8) $M_{\text{HI}}/L_B$ ($M_\odot/L_\odot$) | (9) $S_{\text{H}\alpha}$ ($\frac{10^{-14}\text{erg}}{\text{cm}^2\,\text{s}}$) | (10) $W_{eq}$ (Å) | (11) $\dot{M}$ ($M_\odot$/yr) | (12) $\log(\dot{M}/L_B)$ ($M_\odot/L_\odot$/yr) | (13) $\log(\tau)$ (yr) |
|---|---|---|---|---|---|---|---|---|---|---|---|---|
| 1 | Zw 499.039 | 21.21 ± 0.35 | 21.15 | 5.9 ± 0.3 | 41.4 ± 1.6 | 49 | 0.21 ± 0.08 | | | | | |
|   | UGC 26a | 20.11 ± 0.14 | 20.07 | 1.7 ± 0.1 | 15.4 ± 0.4 | 47 | 0.49 ± 0.09 | 2.6 | 35 | 0.07 | −10.13 | 9.96 |
|   | UGC 26* | 21.19 ± 0.09 | 21.44 | 13 ± 0.3 | 94.6 ± 1.5 | 234 | 0.54 ± 0.05 | | | | | |
|   | UGC 26b | 21.78 ± 0.08 | 21.87 | 5.2 ± 0.1 | 30.6 ± 0.3 | 76 | 0.59 ± 0.11 | 3.0 | 22 | 0.07 | −10.23 | 10.16 |
| 2 | UGC 60* | | | | 104.3 ± 9.3 | 161 | 0.12 ± 0.02 | | | | | |
|   | UGC 60a | | | | | 40 | | | | | | |
|   | UGC 60b | 19.86 ± 0.09 | 20.62 | 3.1 ± 0.1 | 29.4 ± 0.8 | 45 | 0.08 ± 0.02 | 4.6 | 23 | 0.12 | −10.19 | 9.38 |
| 3 | UGC 598a | 21.71 ± 0.21 | 22.19 | 3.8 ± 0.2 | 23.1 ± 0.8 | 63 | 0.83 ± 0.36 | 1.2 | 25 | 0.04 | −10.25 | 10.41 |
|   | UGC 598* | | | | 56 ± 10.0 | | | | | | | |
|   | UGC 598b | 24.27 ± 0.15 | 25.02 | | 11.8 ± 1.0 | 45 | 1.62 ± 0.58 | | | | | |
| 4 | UGC 752* | | | | 119.7 ± 11.2 | 79 | 0.02 ± 0.01 | | | | | |
|   | Zw 502.004 | 20.11 ± 0.05 | 20.35 | 3.7 ± 1.1 | 33.5 ± 6.8 | 59 | 0.22 ± 0.03 | | | | | |
| 5 | UGC 810 | | | | 114.3 ± 9.1 | 162 | 0.23 ± 0.03 | | | | | |
|   | UGC 820* | | | | 150.7 ± 9.1 | 285 | 0.18 ± 0.02 | | | | | |
| 6 | PPHI 0023+34 | 23.47 ± 0.14 | 23.87 | 4.2 ± 0.1 | 11.9 ± 0.2 | 80 | 3.23 ± 0.55 | | | | | |
| 7 | PPHI 0036+34 | 20.65 ± 0.12 | 20.65 | 3.0 ± 0.1 | 24.2 ± 0.4 | 78 | 0.72 ± 0.14 | | | | | |
| 8 | PPHI 0147+31 | 23.40 ± 0.07 | 23.80 | 4.6 ± 0.2 | 13.5 ± 0.5 | 27 | 1.76 ± 0.52 | | | | | |

Notes to TABLE 5:

Col. 2: Target galaxies are denoted by asterisks. UGC 598, the target galaxy of field 3, was not detected in H I: all data related to this galaxy and all optical data related to the cataloged galaxies which were not reobserved by us were taken from the RC3 catalog.

Col. 3: Extrapolated $B$-band central surface brightness, determined by an extrapolation of a linear fit to the straight part of the radial luminosity profile plotted on a magnitude scale. The errors are the $1\sigma$ errors in the $y$-estimate of the regression.

Col. 4: Extrapolated $B$-band central surface brightness, corrected for Galactic and internal absorption following Burstein & Heiles (1984) and Tully & Fouqué (1985) and corrected for line-of-sight integration.

Col. 5: Scale length in seconds of arc, determined by a linear fit to the straight part of the radial luminosity profile plotted on a magnitude scale. At 50 Mpc, $1''$ corresponds to 0.24 kpc. The errors are computed from the $1\sigma$ errors in the estimate of the $x$-coefficient of the regression.

Col. 6: Diameter at $B = 25$ mag arcsec$^{-2}$, in seconds of arc. This was determined by a linear fit to the straight part of the radial luminosity profile plotted on a magnitude scale, except for UGC 598b, in which case an elliptical fit to the $B = 25$ mag arcsec$^{-2}$ isophote was used. The errors are computed from the $1\sigma$ errors in the estimates of $y$ and $x$-coefficients of the regression.

Col. 7: H I diameter, obtained by fitting gaussians to the H I distributions and deconvolving these with the beams.

Col. 8: Total H I mass divided by the blue luminosity. The errors are computed from the $1\sigma$ errors in H I mass and $B$-band luminosity.

Col. 9 and 10: The measured H$\alpha$ flux and equivalent width, respectively.

Col. 11: The star formation rate (SFR) derived as described by Hunter & Gallagher (1986). The measured fluxes were corrected for Galactic absorption.

Col. 12: The logarithm of the ratio of the SFR and blue luminosity, both corrected for Galactic absorption only.

Col. 13: The logarithm of the timescale for gas depletion $\tau$ (in years). This has been computed by dividing the H I mass, multiplied by 1.34 to account for the presence of He, by the SFR.

Table 6: Average optical colors.

| Color | Control sample | | Our sample | | |
|---|---|---|---|---|---|
| | All | Types$\geq$ Sd | All | H I-dwarfs | Cataloged |
| $B-R$ | $1.20 \pm 0.27$ | $1.19 \pm 0.48$ | $0.91 \pm 0.18$ | $0.85 \pm 0.15$ | $1.05 \pm 0.15$ |
| $R-I$ | $0.54 \pm 0.17$ | $0.42 \pm 0.33$ | $0.53 \pm 0.21$ | $0.52 \pm 0.24$ | $0.54 \pm 0.06$ |

Table 7: Group Mass Estimates

| (1) | (2) | (3) | (4) | (5) | (6) | (7) | (8) |
|---|---|---|---|---|---|---|---|
| Field | Name | $\Delta V$ | $\Delta R$ | $M_{\rm est}$ | $M_{\rm min}$ | $M_{\rm est}/L_B$ | $M_{\rm min}/L_B$ |
| | | (km s$^{-1}$) | (kpc) | ($10^{10}\,M_\odot$) | ($10^{10}\,M_\odot$) | ($M_\odot/L_\odot$) | ($M_\odot/L_\odot$) |
| 1 | | | | 177 | | 131 | |
| | Zw 499.039 | $-2.3$ | 306.5 | | 0.02 | | 0.01 |
| | UGC 26a | 42.5 | 167.4 | | 3.5 | | 2.7 |
| | UGC 26b | $-108.9$ | 219.7 | | 30.3 | | 8.9 |
| 2 | | | | 61 | | 15 | |
| | UGC 60a | $-70.7$ | 30.5 | | 1.8 | | 0.5 |
| | UGC 60b | 48.6 | 153.5 | | 4.2 | | 1.2 |
| 3 | | | | 2266 | | 2931 | |
| | UGC 598a | 480.4 | 301.3 | | 807.9 | | 703.5 |
| | UGC 598b | 327.1 | 194.4 | | 241.7 | | 214.3 |
| 4 | | | | 343 | | 191 | |
| | Zw 502.004 | 217.5 | 109.1 | | 59.9 | | 19.5 |
| 5 | | | | 111 | | 50 | |
| | UGC 810 | $-138.2$ | 90.3 | | 20.0 | | 4.9 |

Notes to TABLE 7:

Cols. 3 and 4: Line-of-sight velocity difference and projected distance relative to the central galaxy of the field.

Col. 5: Total mass of the group, computed using the average mass estimator of Heisler et al. (1985).

Col. 6: Minimum pair mass needed to bind a companion to the central galaxy of the field.

Col. 7: Total estimated mass divided by total blue luminosity of the group of galaxies.

Col. 8: Minimum binding mass of each companion-central galaxy pair divided by the total blue luminosity of this pair.